\documentclass[twocolumn,amssymb,showpacs,superscriptaddress,aps,prl]{revtex4-1}
\usepackage{}
\usepackage{stmaryrd}
\usepackage{amsmath}
\usepackage{amssymb}
\usepackage{float}
\usepackage{graphicx}
\usepackage{titletoc}
\usepackage{booktabs}
\usepackage{array, color}
\usepackage{tabularx}
\usepackage{latexsym,bm}
\usepackage{graphics}

\begin{document}

\title{Strong-field ionization inducing multi-electron-hole coherence probed by attosecond pulses}

\author{Jing Zhao}
\affiliation{Department of Physics, College of Science, National University of Defense Technology, Changsha 410073, Hunan, People's Republic of China}
\author{Jianmin Yuan}
\affiliation{Department of Physics, College of Science, National University of Defense Technology, Changsha 410073, Hunan, People's Republic of China}
\affiliation{IFSA Collaborative Innovation Center, Shanghai Jiao Tong University, Shanghai 200240, People's Republic of China}
\author{Zengxiu Zhao}
\email{zhaozengxiu@nudt.edu.cn}
\affiliation{Department of Physics, College of Science, National University of Defense Technology, Changsha 410073, Hunan, People's Republic of China}

\date{\today}

\begin{abstract}
We propose a new scenario to apply IR-pump-XUV-probe schemes to resolving strong field ionization induced and attosecond pulse driven electron-hole  dynamics and coherence in real time.  The coherent driving of both pulses correlates the dynamics of the core-hole and the valence-hole by coupling multiple continua, which leads to the otherwise forbidden absorption and emission of high harmonics. 
An analytical model is developed based on the strong-field approximation by taking into account of the essential multielectron configurations. 
The emission spectra from the core-valence transition and the core-hole recombination are found modulating strongly as functions of the time delay between the two pulses, suggesting the coherent electron wave packets in multiple continua can be utilized to  temporally resolve the core-valence transition in attoseconds.
\end{abstract}

\maketitle

Recent advances in attosecond spectroscopy has enabled resolving electron-hole dynamics in real time \cite{Goulielmakis10nat,ottnature14,McFarland14,Young10,Erk14,schultze10,vrakking14}. The correlated electron-hole dynamics and the resulted coherence are directly related to how fast the ionization is completed \cite{Goulielmakis10nat,ottnature14,Sabbar15}.  Combining the ever-shorter attosecond pulses with intense infrared lasers, it is possible to probe and control both core and valence electrons coherently on the equal footing. 
However, it is sitll challenging to answer these key questions in attosecond physics or even attosecond chemistry \cite{Leone14, Golubev15} such as how the coherence evolves  and transfers among electron-hole pairs  or multiple channels.

Coherent X-ray pulses with duration of femtosecond or attoseconds have been generated from free-electron lasers \cite{Emma10,Rohringer12} or high-harmonic generation (HHG)  \cite{Paul01,hentschel01,zhao67as}.  They are capable of creating inner holes followed with exotic correlated electron dynamics  such as cascading Auger processes \cite{Young10,Erk14,McFarland14,Li15} and ionization induced  absorption saturation \cite{Rackstraw15,fukuzawa13}.  
On the other hand, intense infrared (IR) lasers ignite ionization from the valence shell. The released  electron and the created hole are still driven by the external fields. When the field reverse its direction, the electron recombines into the original hole  leading to HHG. However, due to the multielectron nature,  many-body effects, e.g., dynamical polarization \cite{zhang13prl, zhao14pra,bauer14pra} will modify the dynamics of both the electron and the hole.  Particularly, previous work has demonstrated that multielectron information is encoded in HHG  when the inner shell electrons are excited by the IR lasers and participating in the ultrafast dynamics \cite{Zhaoprl06,Santra06L,santra13,zhang14pra,leeuwenburgh2013,leeuwenburgh2014}, but the effect on HHG from coherent driving of core-valence transition by attosecond pulses remains unexplored. 

\begin{figure}
\includegraphics[width=3.0in]{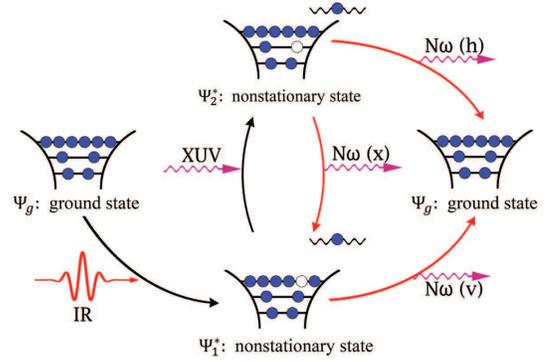}
\caption{\label{corevalencehhg} Illustration of the interaction of the atom with the laser and XUV fields and  the related high-harmonic generation processes: Ionization from the filled valence shell  by the laser field creates the nonstationary state $\Psi_1^*$ with its associated continuum $\phi_v$. Concurrently, it opens the subshell allowing the followed resonant transition pumped by the XUV pulse, which creates a hole in the core and  transfers the continuum into $\phi_h$, i.e., to the nonstationary state $\Psi_2^*$. Recombination into the valence shell (path (v)) and the core hole (path (h)) could emit high-order harmonics. The resonant transition from the two nonstationary states (path (x)) also generate coherent harmonics.}
\end{figure}

In this Letter, we consider a unique scenario involved atoms with closed shells subjected to an intense IR laser pulse and a time-delayed attosecond pulse (AP) which has a central frequency in resonant with the transition between the inner and valence shells. In the absence of the IR pulse, the direct transition from the inner shell to the valence shell is forbidden due to the Pauli exclusion principle. 
However, once the IR field induces ionization from the valence shell, the transition is triggered leaving a hole in the inner shell affecting the subsequent rearrangement dynamics.  As shown in Fig.~\ref{corevalencehhg}, strong field ionization from the filled valence shell  by IR field creates the associated continuum. Concurrently, it opens the subshell allowing the followed resonant transition pumped by the AP, which creates a hole in the core and  transfers the continuum into its own.
The attosecond light absorption and the resulted emission are thus gated by the ionization, in close analogy to the ionization induced absorption saturation where the transition energy is shifted by ionization \cite{fukuzawa13}. 
Meanwhile, the opened AP absorption creates coherence between the valence-hole and the core-hole.  
The transfer of coherence from ionization into both holes leads to multiple paths of HHG: harmonics can be radiated through recombination into the valence shell (path (v)) and the core hole (path (h)) respectively, or it can be generated  upon the resonant core-valence transition accompanied by the transfer of the continua (path (x)). 
 The coherent electron wave packet in multiple continua thus provides the opportunity to temporally resolve the multi-electron-hole dynamics in attoseconds. 

The essential multielectron configurations involved are the neutral ground state $\Psi_g$, the non-stationary states 
$\Psi_1^*$ which constitute of  the ionic ground state $\Phi'_1$ with the $n$-th electron in the continuum $\psi_v({\bf r_n},t)$, and $\Psi_2^*$ with the ionic core in excited state $\Phi'_2$ and the released electron in the continuum $\psi_h({\bf r}_n,t)$.
% and the excited state $\Phi'_2$ of  with the released electron in the respective continuum, $\psi_v({\bf r}_n,t)$ and $\psi_h({\bf r}_n,t)$ respectively. Here the $i$-th electron coordinates are denoted by ${\bf r}_{i}$. For simplicity, we avoid explicitly writing out the ($n-1$) electron coordinates and  replace ${\bf r}_n$ by ${\bf r}$ in the following. 
  Taking Ne atom as an example, we have $\Phi'_1=\Phi_{1s^22s^22p^5}$ and $\Phi'_2=\Phi_{1s^22s^12p^6}$ which can be formed by ionization from the valence shell or inner shell respectively, while the latter  resembles the situation that a hole is created in the inner shell. 
The total time-dependent wave function of the $n$-electron atom subjected to external laser fields can thus be approximated as 
\begin{align}
\Psi(t)&=a_g(t)\Psi_ge^{-iE_gt}+\hat{\mathcal{A}}\left[\Phi'_1\psi_{v}(t)\right]e^{-iE'_1t}\nonumber\\
&+\hat{\mathcal{A}}\left[\Phi'_2\psi_{h}(t)\right]e^{-iE'_2t},
\end{align}
where $\hat{\mathcal A}$ is the antisymmetrizing operator on the electron coordinates.  
%where $\hat{\mathcal A}\,=\,\left(1-\sum_{i=1}^{n-1}\hat{P}_{in}\right)/\sqrt{n}$ is the antisymmetrizing operator on the electron coordinates.
$E_g$, $E'_1$ and $E'_2$ are the binding energies of the neutral ground state, the ionic ground and excited states respectively.   Assuming that the three configurations are orthogonal, the probability of finding the ion in the ground or excited states are given by the normalization of the continuum electrons, denoted by $||\psi_v||^2$ and $||\psi_h||^2$ respectively. The probability amplitude of the atom remaining in the neutral ground state is denoted by $a_g$ and    $|a_g|^2+||\psi_v||^2+||\psi_h||^2=1$.

From the time-dependent Schr\"{o}dinger equation in a linearly polarized electromagnetic field ${\bf E}(t)$, 
we obtain the coupled-channel equations for the $n$-th electron,
\begin{eqnarray}
i\dot{\psi_v}=\tilde{H_v}\psi_v+a_g(t)\langle\Phi'_1|{\bf r}\cdot{\bf E}(t)|\Psi_g\rangle' e^{iI_1t}\nonumber\\
+\langle\Phi'_1|\sum_{i=1}^{n-1}{\bf r}_i\cdot{\bf E}(t)|\Phi'_2\rangle' \psi_he^{-i\Delta It}, \\
i\dot{\psi_h}=\tilde{H_h}\psi_h+a_g(t)\langle\Phi'_2|{\bf r}\cdot{\bf E}(t)|\Psi_g\rangle' e^{iI_2t}\nonumber\\
+\langle\Phi'_2|\sum_{i=1}^{n-1}{\bf r}_i\cdot{\bf E}(t)|\Phi'_1\rangle' \psi_ve^{i\Delta It},
\end{eqnarray}
where $\langle |...|\rangle'$ denotes the integration over the ($n-1$)-electron coordinates, $I_{1,2}=E'_{1,2}-E_g$ and $\Delta I=I_2-I_1$.
$\tilde{H}_{v,h}$ are the effective hamiltonians for the excited electron in the laser field with 
the core left in  the two ionic states respectively. For simplicity, we have replaced ${\bf r}_n$ by ${\bf r}$.
The second terms in Eq.~(2) and (3) originate from the transition from the neutral ground state to the two ionic states by prompting the $n$-th electron into their respective continuum, while the third terms represent the couplings between the two ionic channels induced by the external fields. 

The central frequency of the XUV pulse is chosen to be exactly matching the transition energy $\Delta I$ from the inner shell to the valence shell. 
The total electric field is given by  ${\bf E}(t)={\bf E}_{\rm L}(t)+{\bf E}_{\rm X}(t)$, where the IR laser ${\bf E}_{\rm L}(t)$  has a frequency far less than $\Delta I$ and  therefore its contribution to the core-valence  transition is negligible in the absence of multiphoton resonant excitation. 
For ionization from the valence shell (Eq.~(2)), we include only direct ionization from the neutral ground state by the IR field, while neglecting  the coupling from the core excitation induced by the AP.

Denoting the respective Green's functions of $\tilde{H_v}$ and $\tilde{H_h}$ as $\tilde{G_v}$ and $\tilde{G_h}$, we obtain
\begin{eqnarray}
\psi_v(t)=\int^tdt'\tilde{G}_v(t,t')a_g(t'){\bf E}_{\rm L}(t') e^{iI_1t'}\varphi',\\
\psi_h(t)=\int^tdt'\tilde{G}_h(t,t'){\bf d}^*_{12}{\bf E}_{\rm X}(t') e^{i\Delta It'}\psi_v(t').
\label{cvtran}
\end{eqnarray}
The  core-valence transition dipole moment between the two ionic states $\Phi'_1$ and $\Phi'_2$ is given by ${\bf d}_{12}\,=\,(n-1)\langle\Phi'_1|\sum_{i=1}^{n-1}{\bf r}_i|\Phi'_2\rangle'$ and 
$\varphi'\,=\,\langle\Phi'_1|{\bf r}|\Psi_g\rangle'$ is the valence-associated Dyson orbital weighted by the dipole operator.
In reality, there should be difference between the Green's functions $\tilde{G}_v$ and $\tilde{G}_h$  because of the core-electron rearrangement. For example, the potential felt by the continuum electron varies when the core making transition from the valence shell to the inner shell. The difference is ignored in the present model since  the atomic potential on the continuum electron is ignored comparing to the laser field as in the strong field approximation \cite{Lewenstein94}.
Therefore the Green's function can be written in term of Volkov states
\begin{equation}
%\tilde{G}_{h,v}(t,t')=-i\int d{\bf p}\exp\lbrack-iS({\bf p},t,t')\rbrack|{\bf v}(t)\rangle\langle{\bf v}(t')| \label{volgf},
\tilde{G}_{h,v}(t,t')=-i\int d{\bf p}e^{-iS({\bf p},t,t')}|{\bf p}+{\bf A}(t)\rangle\langle{\bf p}+{\bf A}(t')|, \label{volgf}
\end{equation}
with the semiclassical action $S({\bf p},t,t')=\int_{t'}^t\frac{({\bf p}+{\bf A}(\tau))^2}{2}d\tau$ and the vector potential of the field ${\bf A}(t)=-\int^t{\bf E}(t)dt$.

%The harmonic emission spectrum is obtained from the Fourier transformation of the induced dipole moment along the laser polarization,
%which can be written as ${\bf d}(t)\,=\,{\bf d}_v(t)\,+\,{\bf d}_h(t)\,+{\bf d}_{x}(t)$ with
The time-dependent dipole moment can be divided into three parts
\begin{align}
{\bf d}_v(t)\,=\,a^*_g(t)e^{-iI_1t}\left(\langle\psi_1^D|{\bf r}|\psi_v\rangle\,+\,\langle{\bf \varphi}_1|\psi_v\rangle\right)\,+\,c.c,\label{eqdv}\\
{\bf d}_h(t)\,=\,a^*_g(t)e^{-iI_2t}\left(\langle\psi_2^D|{\bf r}|\psi_h\rangle\,+\,\langle{\bf \varphi}_2|\psi_h\rangle\right)\,+\,c.c,\label{eqdh}\\
{\bf d}_x(t)\,=\,e^{-i\Delta It}\left(   \,{\bf d}_{12}\langle\psi_v|\psi_h\rangle\,+{\bf d}_{vh}\,+\,{\bf d}_{ex}\right)\,+\,c.c.\label{eqdex}
\end{align}
Here ${\bf d}_v(t)$ and ${\bf d}_h(t)$ are related to the paths (v) and (h) of HHG illustrated in Fig.~\ref{corevalencehhg} respectively. They both consist of an effective one-electron transition with the ionization channel-specific
Dyson orbital $\psi_{1,2}^D\,=\,\sqrt{n}\langle\Phi'_{1,2}|\Psi_g\rangle'$, and an exchange correction term
as introduced in \cite{Zhaoprl06} with ${\bf \varphi}_{1,2}\,=\,\sqrt{n}\langle\Phi'_{1,2}|\sum_{i=1}^{n-1}{\bf r}_i|\Psi_g\rangle'$, which arises from indistinguishable of electrons. The path (x) of HHG is determined by the dipole moment ${\bf d}_x(t)$ from the transition between the two non-stationary states.

%
%This multi-electron collision plays a crucial role in strong-field ionization in the form of dynamical core polarization \cite{zhang13prl,zhao14pra,zhang14pra,bauer14pra}, and manifest itself in high-harmonic generation \cite{Zhaoprl06,santra13,zhang14pra}.

\begin{figure}
\includegraphics*[width=3.0in]{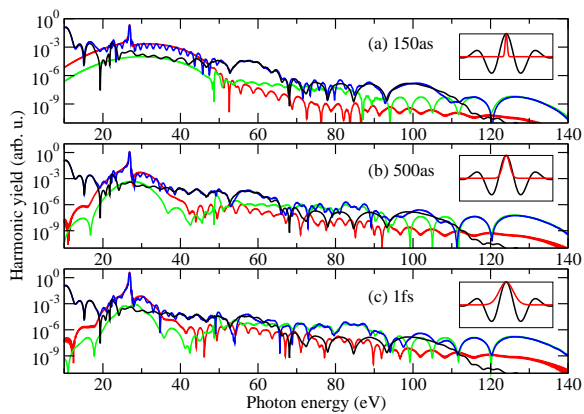}% Here is how to import EPS art
\caption{\label{hgspectra} Harmonic spectra calculated from  ${\bf d}_v$ (black), ${\bf d}_{x}$ (red), ${\bf d}_h$ (green) and the total dipole moment ${\bf d}(t)$ (blue), at the time delay of 0 between the two pulses, at different XUV pulse durations of 150\,as, 500\,as and 1\,fs from (a) to (c). The electric field and the envelop of the APs are shown in the insets. The laser intensity is 4$\times$10$^{14}$\,W/cm$^2$ and the XUV pulse intensity is 1$\times$10$^{13}$\,W/cm$^2$ for all cases.}
\end{figure}

To mimic Ne atom, we choose the ionization potential of the  valence shell as $I_1=21.56$\,eV 
and the transition energy of $\Delta I=26.89$\,eV corresponding to $2s$ to $2p$ of Ne atom.  
In Fig.~\ref{hgspectra}, the emission spectra calculated from the three individual processes 
are presented, as well as the total spectra, at the time delay of 0.
The IR field has a Gaussian envelop with one-cycle duration and a central wavelength of 800\,nm. 
The spectra directly calculated from ${\bf d}_v$, the recombination into the valence shell, 
are the same as that obtained without XUV pulse which has been assumed to induce core-valence transition only. 
%Certainly the XUV pulse could directly ionize electrons from the valence shell of the neutral, as shown in \cite{Johnsson05} when its photon energy is larger than the ionization potential. Or it can induce ionization modulating with the delay between the XUV pulse and the IR pulse when the XUV photon energy is small but close to the ionization potential \cite{Johnsson07}.

In Fig.~\ref{hgspectra}, it can be seen that the cutoff of the total spectra (blue lines) is extended comparing to 
that emitted through path (v) (black lines).   Once the hole in the inner shell is created, the electron can directly recombine into the hole through path (h) (green lines) and  harmonics are emitted 
%with photon energy given by $I_2+E_k$ with $E_k$ being the kinetic energy of the electron. 
%Thus 
with the cut-off energy extended by $\Delta I$,
which has also been observed in HHG from multiple orbitals in molecules with ionization from one orbital and recombination to a lower-lying orbital, either coherent driven by strong IR field \cite{mairesse10prl} or magnetic field \cite{cireasa2015np}. 

For harmonics generated from path (x) (red lines),
a pronounced peak appears at the core-valence transition energy.
The total emission of harmonics (blue lines) around the resonance peak is enhanced by orders and therefore dominated by the path (x). 
%The harmonic yield around the resonance peak in the total emission is therefore enhanced by orders.
As the XUV pulse duration becomes shorter, the continuum background of the spectra become broader,
however, the profile of the peak remains the same. In the simulation, we include only the first term in Eq.~\ref{eqdex} that originates from the bound-bound transition among the two ionic states  while the continuum makes jumping. 
The second term ${\bf d}_{vh}\,=\,\langle\Phi'_1|\Phi'_2\rangle'\langle\psi_v|{\bf r}|\psi_h\rangle$ arises from the continuum-continuum transition among $\psi_v(t)$ and $\psi_h(t)$ weighted by the overlap of two ionic states.
Because it usually gives rise to low frequency emission, we neglect it here. 
The third term  ${\bf d}_{ex}\,=\,2\sum_{i=1}^{n-1}\langle\hat{P}_{in}\Phi'_1\psi_v|{\bf r}|\Phi'_2\psi_h\rangle$ describes a two-step process as in Ref.~\cite{santra13},
where the electron released from the valence orbital, promotes the inner-shell electron to the vacancy it created upon recollision, and recombines into the newly formed hole to emit harmonics. Comparing to the XUV resonant excitation in this study, the probability of collision-induced rearrangement is much smaller and therefore this term is neglected as well.

Note that the spectrum is not simply reflecting the line-shape of the core-valence transition, as seen by the higher energy photon of emission. 
The path (x) is in fact  from the coherent transition between two non-stationary states $\Psi^*_1$ and $\Psi^*_2$, similar to autoionization states with one electron embed in the continuum interacting loosely with the ionic core. 
Because the driving of  the external laser fields, the total energy of the non-stationary state is varying with time and spreading over a broad range due to the correlation between the continuum electron and the ion.  Their energy difference gives rise to the emission of higher energy photon whose yields are determined by both the transition dipole of the ionic states and  the { \it temporal correlation between the two associated continua}.
When the contributions from all the three paths are comparable, their interference produces  a generalized Fano profile (see the total emission around 70eV), similarly to the laser-assisted autoionization \cite{Zhao05a,Wickenhause05,Zielinski15} where the autoionization profile is evolving by the laser fields.
\begin{figure}
\includegraphics*[width=3.0in]{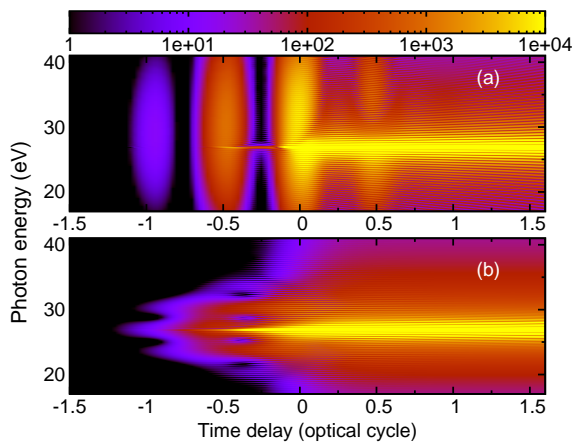}% Here is how to import EPS art
\caption{\label{corevalence} Variation of the emission spectra induced by core-valence transition at different time delay between the IR pulse and APs with duration of (a) 150 as and (b) 1fs. }
\end{figure}

By varying the time delay between the IR and the XUV pulses, the emission spectra from ${\bf d}_x$ changes as shown in Fig.~\ref{corevalence} for the XUV pulse duration of 150 as and 1 fs.   
 The spectra intensity is related to the vacancy in the valence shell induced by tunnelling ionization. 
The sensitivity of  ionization to the instantaneous field, leading to the strongly modulated delay-dependent emission spectra in Fig.~\ref{corevalence}. Especially when the AP comes before the maximum of the IR field (negative delay), the ionization probability is small and  the emission is very weak as the emission  from core-valence transition is prohibited without ionization of the valence shell.  
%The fringe pattern appearing after zero time delay in Fig.~\ref{corevalence} (a) originates from the interference between the XUV pulse and the emitted harmonics, which has also been observed in delay-dependent transient absorption spectra and photoelectron spectra \cite{ottnature14,jing12}. 

\begin{figure}
\includegraphics*[width=3.0in]{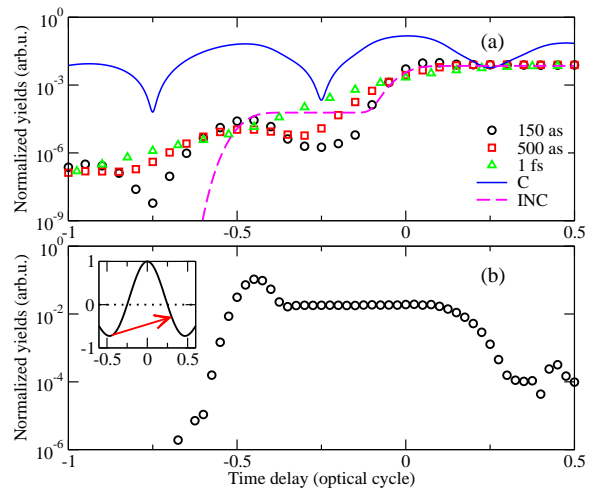}% 
\caption{\label{iondyn} (a)Harmonic emission yields integrated from 25.89\,eV to 27.89\,eV with AP durations of 150\,as (circle), 500\,as (square) and 1\,fs (upward triangle). Dashed and dash-dot lines represent the ionization probability obtained  from incoherent integration of the ionization rate at each instantaneous field and from $||\psi_v||^2$ respectively. (b) Integrated emission yields over the cutoff region from 120\,eV to 130\,eV versus the time delay with AP duration of 150\,as. The arrow in the inset points out ionization and emission istants of the cut-off harmonic.}
\end{figure}

In order to quantify the varying of emission with respect to the time delay, we integrate the total harmonic yields around the resonance peak over [25.89,27.89] eV shown in Fig.~\ref{iondyn} (a) for a couple of different AP durations.  The yields reaches a local maximum whenever the laser field strength passes a local maximum.  This  is related to the fact that the ionization rate peaks at those instants and leaves more vacancy in the valence shell. 

Interestingly, the yields exhibit modulation with time delay, which is against the  expectation that the ionization probability is increasing monochromatically when incoherent integrating the ionization rate over time (dashed lines in Fig.~\ref{iondyn} (a)).  
 On the other hand, the ionization probability obtained from $||\psi_v||^2$ at each instant is exhibiting modulation  (solid lines in Fig.~\ref{iondyn} (a)), because the coherent electron wavepackets generated at different instants interference with each other,  leading to the modulation of the population of vacancy.  For comparison, we have normalized the ionization probability at the end of the pulse obtained from the two approaches.  
 The coherence of electron wavepacket within ionization is thus imprinted on the sum yields since which are related to the temporal correlation of the electron wavepacket as stated above. The longer the AP is, the more averaged vacancy is probed and the time-delayed sum yields are less contrasted. As shown in Fig.~\ref{iondyn} (a), the emission yield turns into almost a smooth line for AP duration of 1 fs.

More than probing the coherence of the valence-hole, the AP probes the coherence of the core-hole as well. During the  propagation, the electron jumps between the two ionic state-associated continua because of the driving of the AP. An additional phase shift  will be induced from this channel-coupling.  Although it is neglected so far by assuming $\tilde{G}_h\approx \tilde{G}_v$ in Eq. \ref{volgf}, we expect that it will be evident in the sum yields since which measures the  time correlation between the two continuum wave packets. 

The coherence information of the core-hole is also encoded in harmonic spectra near the extended cut-off region contributed by the HHG path (h). Because of the applied few-cycle laser pulse, the cut-off harmonics are mainly 
 contributed by the electron born half cycle before the envelop peak and recombines into the core-hole around 0.2 optical cycle past the peak (see the inset of Fig.~\ref{iondyn} (b)), on the condition that the core-hole has been formed  within this time interval.   The sum yields within [120eV, 130eV] interval are  found exhibiting a platform when the AP is applied within this interval.  The plateau starts with a burst at -0.5 cycle  which is caused by the higher rate of coherently creating the core-hole in the combination of the IR field and the AP.  
While the width of the burst reflects the temporal behavior of the core-valence transition driven by both pulses, the width of the plateau is determined by the duration of recombination if the spontaneous decay or other relaxation processes of the core hole are much longer.

In conclusion, we propose an IR-pump-XUV-probe scheme to investigate the interplay of the valence-hole and the core-hole created from atoms with filled valence shells. Using the laser-induced ionization as a gate for XUV excitation of core electrons, it provides us
the opportunity to probe both the core and valence electron dynamics by manifesting themselves as a pronounced resonant peak in harmonic spectra and an extended cut-off harmonic emission. The coherent interplay of multiple paths of HHG is found evident in the profile of the harmonic spectra. By analyzing the modulation of the spectra with the time-delay between the IR field and the AP,  we show that the coherence of the ionization process and the driven core-hole and valence-hole coherence contribute to HHG which can be utilized to obtain the multi-electron-hole or multichannel coherent information.

This work is supported by the National Basic Research Program of China (973 Program) under Grant No. 2013CB922203, and the National Natural Science Foundation of China (Grants No. 11374366 and 11404401). We acknowledge valuable discussions with Yongqiang Li, Xiaowei Wang and Yindong Huang.
%
%\bibliography{xuvhhg}
%\end{document}

%

\end{document}